\documentstyle[epsf,prd,aps]{revtex}

\newcommand{\gamo}{\gamma^0}
\newcommand{\gamr}{\gamma^{r}}
\newcommand{\gamt}{\gamma^\theta}
\newcommand{\gamtr}{\gamma^3}
\newcommand{\derir}{\partial_{r}}
\newcommand{\derit}{\partial_\theta}
\newcommand{\deriz}{\partial_z}
\newcommand{\ud}{{\mathrm{d}}}

\def\beq{\begin{eqnarray}}
\def\eeq{\end{eqnarray}}
\newcommand{\nn}{\nonumber}
\newcommand{\reals}{\mbox{${\rm I\!R }$}}

\title{Massive 3+1 Aharonov-Bohm fermions in an MIT cylinder }
\author{M. De~Francia and K. Kirsten}
\address{Department of Physics and Astronomy, The University of Manchester, \\
Theory Group, Schuster Laboratory, Manchester M13 9PL, England}
\date{\today}

\begin{document}
\bibliographystyle{prsty}
\maketitle
\begin{abstract}
We study the effect of a background flux string on the vacuum energy of
massive Dirac fermions in 3+1 dimensions confined to a finite spatial
region through MIT boundary conditions. We treat two admissible
self-adjoint extensions of the Hamiltonian. 
The external sector is also studied  
and unambiguous results for the 
Casimir energy of massive fermions in the whole space are obtained.

PACS number(s): 11.15.Kc, 12.39.Ba, 02.30.-f
\end{abstract}


\section{Introduction}
The influence of background fields or of boundaries on the vacuum structure
of a quantum field theory is of continuing 
importance in various branches
of modern physics. The various 
aspects are often described generically as Casimir effects and 
many techniques have been developed in order to
analyse the different situations. As examples we mention the
Green's function approach
\cite{milt99pre,schw78-115-1,brev94-27-6853} and the zeta function
regularization
\cite{hawk77-55-133,dowk76-13-224,eliz94b,eliz95b}.

In the years since the Casimir effect was first discussed \cite{casi48-51-793}
these techniques have been refined considerably. 
Relatively recently, a contour
integral method has been developed which allows the representation of the
Casimir or ground state energy in terms of eigenfunctions or of the
Jost function of the associated field equation
\cite{bord96-53-5753,bord96-37-895,lese96-250-448}. This allows for a relatively
direct analysis for configurations where this information can be easily obtained. As a result, spherically or cylindrically symmetric situations have
been analysed in various contexts, see
e.g.~\cite{lese96-250-448,bord99-59-085011,bord97-56-4896,gosd98-441-265,milt99-59-105009,bene00-61-085019,scan00-33-5707,scan99-32-5679,scan00-62-085024}.
Most of the research done so far concentrates on the influence of background
fields or of boundaries separately. Relatively little is known when these two
effects are combined. Exceptions are
e.g.~\cite{lese98-193-317,bene00-61-085019}, where a magnetic fluxon and
MIT boundary conditions for a massless and massive Dirac field in
$2+1$ dimensions were considered.

It is our aim to continue the analysis of this combined effect in a $3+1$
dimensional cylindrically symmetric situation. We consider a massive Dirac
operator in the presence of a magnetic flux string located along the
$z$-axis and with MIT boundary conditions
at a cylinder of radius $R$. Because of the presence
of the flux string, a self-adjoint extension of the radial Dirac
operator is needed, as it is well known from the $2+1$ dimensional situation
\cite{gerb89-124-229,gerb89-40-1346}. In analogy to
\cite{bene00-61-085019} we will consider two possible self-adjoint
extensions which are both compatible with the presence of a Dirac delta
magnetic field at the origin. One extension is constructed by imposing spectral
boundary conditions \cite{atiy75-77-43,atiy75-78-405,atiy76-9-71} at a finite
radius which is 
then shrunk to zero. The second one follows from the zero radius limit
of a cylindrical flux shell \cite{alfo89-328-140,hage90-64-503,flek91-6-5327}.
The Casimir energies differ, reflecting the fact that different
self-adjoint extensions describe different physics in the core.

Compared to the calculations in $2+1$ dimensions, peculiar differences
occur. Most importantly, when considering 
only the interior of the cylinder,   
the poles in the Casimir energy depend on the flux
for both extensions. 
This renders an interpretation of the results 
impossible, in the sense that finite values cannot be extracted in a 
physically reasonable way. This led us to consider also 
the exterior space, in an attempt to get a simpler pole structure by 
adding up both contributions. In doing so, 
unambiguous results for Casimir energies in different situations 
can be obtained. Let us stress that in the present situation 
general heat-kernel arguments \cite{blau88-310-163,dowk84-1-359} 
can not be applied simply because no general answers are known for 
cases with singular fluxes and boundaries. For this reason we really 
need to perform explicit calculations.

The paper is organized as follows. First we briefly present the
solutions of the Dirac equation in the presence of a flux string. In the
following sections self-adjoint extensions are obtained and 
some details of how
to impose spectral boundary conditions are provided. Having chosen the self-adjoint extensions, the MIT boundary conditions are imposed and implicit
eigenvalue equations are obtained. These are the basis for the analysis
of the Casimir energy. In order to obtain a well-defined Casimir energy
(such that a numerical analysis makes sense) we have evaluated 
differences between the case when the flux is arbitrary and when the 
flux is integral. Also, we include the exterior space and the 
pertinent changes in the calculation are briefly explained. A numerical 
analysis of Casimir energies as a function
of the flux, for the Dirac field, is presented
afterwards.

\section{Description of the problem}

We study the Dirac equation for a massive particle in 3+1 dimensional 
Minkowski space-time in the presence of a flux string located at the origin,
\beq
\vec H = \nabla \wedge \vec A = \frac{\kappa} r \delta (r) \vec e_z .
\eeq
We are going to impose boundary conditions on a cylinder and for that 
reason use cylindrical coordinates. The gamma matrices in the chiral 
representation are 

\[
\gamo = \rho_3 \otimes \sigma_3, \quad 
\gamma^1= i \rho_3 \otimes \sigma_2,;
\quad \gamma^2 = - i \rho_3 \otimes \sigma_1,; 
\quad \gamma^3 = i \rho_2 \otimes
\mathrm{1}_2,
\]
and in a comoving coordinate frame one has the usual relation
\[
\gamr = \cos \theta \gamma^1 + \sin \theta \gamma^2, \quad
\gamt = \frac1r \left( - \sin \theta \gamma^1 + \cos \theta \gamma^2
\right).
\]
Furthermore, for the flux
\[
A^\theta = \frac{\kappa}{r}, \quad \kappa = l + a, \quad 0 \leq a <1.
\]
Here we have introduced the integer part $l$ of the reduced flux 
$\kappa$ such that $a$ is restricted to the given interval.

Our main concern is the Casimir energy and for this reason we consider
the Dirac Hamiltonian for the system,
\begin{equation}
H_{\mathrm D} \Psi_E= -i \gamo \gamr \left[ - \derir + \gamr \gamt (\derit - i \kappa) +
\gamr \gamtr \deriz + i m \gamr \right] \Psi_E = E \Psi_E.\label{eq:1}
\end{equation}
The representation of the Hamiltonian is chosen in such in a way as 
to simplify 
the implementation of spectral boundary conditions at a later stage.
Given the cylindrical symmetry of the configuration the eigenfunctions
have the form
\begin{equation}
\Psi_E (r,\theta,z) =  e^{-i k_z z}
\psi_e (r, \theta), \quad k_z \in \reals.
\end{equation}
For $\psi_e (r, \theta)$ there are four types of solutions. They are
\beq
\psi_e^{(n,1)} = \left(
\begin{array}{c}
J_{\omega} (k r) e^{i n \theta} \\
- C_- J_{\omega+1} (kr) e^{i(n+1) \theta} \\
\Gamma J_{\omega} (kr) e^{i n \theta} \\
\Gamma C_- J_{\omega+1} (kr) e^{i(n+1) \theta}
\end{array}
\right),
\psi_e^{(n,2)}  = \left(
\begin{array}{c}
J_{-\omega} (k r) e^{i n \theta} \\
C_- J_{-\omega-1} (kr) e^{i(n+1) \theta} \\
\Gamma J_{-\omega} (kr) e^{i n \theta} \\
-\Gamma C_- J_{-\omega-1} (kr) e^{i(n+1) \theta}
\end{array}
\right),    \label{eq:2}\\
\psi_e^{(n,3)} = \left(
\begin{array}{c}
\Gamma J_{\omega} (k r) e^{i n \theta} \\
\Gamma C_+ J_{\omega+1} (kr) e^{i(n+1) \theta} \\
J_{\omega} (kr) e^{i n \theta} \\
-C_+ J_{\omega+1} (kr) e^{i(n+1) \theta}
\end{array}
\right),
\psi_e^{(n,4)} = \left(
\begin{array}{c}
\Gamma J_{-\omega} (k r) e^{i n \theta} \\
- \Gamma C_+ J_{-\omega-1} (kr) e^{i(n+1) \theta} \\
J_{-\omega} (kr) e^{i n \theta} \\
C_+ J_{-\omega-1} (kr) e^{i(n+1) \theta}
\end{array}
\right),      \label{eq:3}
\eeq
where $n=-\infty,...,\infty$, and we have used the notation 
\[
k=\sqrt{e^2-m^2}, \quad E= \textrm{sign}(e) \sqrt{k_z^2+e^2},
\]
\[
C_\pm = \frac{ik}{e\pm m}, \quad \Gamma = - \frac1{k_z} \left(e -
E \right), \quad \omega = n - l - a.
\]
Here $e^2=k^2+m^2$ are the eigenvalues of the square of the 
two-dimensional Dirac operator obtained after setting $k_z =0$.

\bigskip 

In order to fix the eigenvalues $E$ we have to choose a self-adjoint extension
of the radial Hamiltonian and we have to 
impose boundary conditions. A family of self-adjoint 
extensions arises and we will consider two particular cases.
The first one is obtained by imposing spectral boundary conditions at some
interior cylinder the radius of which is sent to zero. Spectral boundary 
conditions are imposed as described in 
\cite{atiy75-77-43,atiy75-78-405,atiy76-9-71}, see also 
\cite{gilk00pre}. 
In the present case a suitable choice for the boundary operator $A$ is 
\cite{gilk00pre,dowk99-242-107}
\begin{equation}
A = \gamr \gamt (\derit - i \kappa) + \gamr \gamtr \deriz + \frac1{2
r} \mathrm{1}_4.
\end{equation}
The term $(1/2r) \mathrm{1}_4 = (1/2r) K \mathrm{1}_4$, with $K$ the 
extrinsic curvature of the boundary, has been added to guarantee that 
the operator $A$ leads to a self-adjoint boundary value problem, 
see \cite{gilk00pre} for details.

Spectral boundary conditions amount to setting to zero the 
projection of $\Psi_E$ onto all eigenfunctions of $A$ with negative 
eigenvalues at the boundary. 
Again, due to the symmetry, the eigenfunctions have the form
\begin{equation}
A=  e^{-i k_z z}
\alpha,
\end{equation}
where again four different types of solutions exist,
\[
\alpha_1 = \left(
\begin{array}{c}
e^{i n \theta} \\
0 \\
0 \\
d_+ \, e^{i (n+1) \theta}
\end{array}
\right),
\alpha_3 = \left(
\begin{array}{c}
0\\
d_+ \, e^{i (n+1) \theta} \\
e^{i n \theta}\\
0
\end{array}
\right),
\]

\[
\alpha_2 = \left(
\begin{array}{c}
0\\
e^{i n \theta} \\
d_- \, e^{i (n-1) \theta}\\
0
\end{array}
\right),
\alpha_4 = \left(
\begin{array}{c}
d_- \, e^{i (n-1) \theta} \\
0 \\
0 \\
e^{i n \theta}
\end{array}
\right).
\]
Here we have introduced 
\[
d_\pm= \frac{i}{k_z r_0} \left[ a_\pm -\left(\omega \pm \frac12 \right) 
\right], \quad a_\pm =
\textrm{sg} \left(\omega \pm \frac12 \right) \sqrt{\left( k_z r_0 \right)^2 +
\left(\omega \pm \frac12 \right)^2}
\]
and $\alpha_1$, $\alpha_3$ can be seen to have eigenvalues $a_{1,3}=(1/r_0)
a_+$ and $\alpha_2$, $\alpha_4$ have instead $a_{1,4}=-(1/r_0)a_-$.
The eigenvectors and eigenvalues of $A$ in $2+1$-dimensions are 
obtained by taking $k_z \rightarrow 0$.

Having these eigenfunctions $\alpha$ at our disposal, projections with 
$\Psi_E$ can easily be performed. Due care has to be taken in order to 
implement the boundary conditions correctly for all values of $n$. In the 
limit $r_0 \to 0$, the vanishing of the projection rules out some of the 
$\psi_e ^ {(n,i)}$ in Eqs.~(\ref{eq:2}) and (\ref{eq:3}) because
they diverge at the origin. In detail, one finds that for 
$n\geq l+1$ the function $\psi_e^{(n,1)}$ and $\psi_e^{(n,3)}$ 
form a suitable basis, whereas for $n<l$ the relevant ones are
$\psi_e^{(n,2)}$ and $\psi_e^{(n,4)}$.  The remaining one, $n=l$, allows the
identification as  
a member of the one parameter family of self-adjoint 
extensions \cite{gerb89-40-1346}. Writing the spinors in the form 
\beq
\psi_e^ n (r,\theta ) = \left(
\begin{array}{c}
f_n^ + (r) e^{in\theta} \\
g_n^+ (r) e^{i(n+1)\theta} \\
f_n^- (r) e^{in\theta} \\
g_n ^- (r) e^{i(n+1) \theta} 
\end{array}
\right), 
\eeq
self-adjoint extensions are classified according to the conditions
\begin{equation}
i \lim_{r_0 \rightarrow 0} 
(m r_0)^{1-a} g_l^\pm (r_0) \sin \left( \frac{\pi}4 + \frac{\Theta^\pm}2
\right) = \lim_{r_0 \rightarrow 0} 
(m r_0)^{a} f_l^\pm (r_0) \cos \left( \frac{\pi}4 + \frac{\Theta^\pm}2
\right).\label{eq:4}
\end{equation}
In our case, 
for $0<a<1/2$,
\begin{eqnarray}
f_l^+ (r_0) & = &\left[ A_l^+ + \Gamma A_l^- \right] J_{-a} (k r_0) ,
\nonumber \\
g_l^+ (r_0) & = & \left[ - A_l^+ C_- + \Gamma A_l^- C_+ \right] 
J_{-a+1} (k r_0) ,\nonumber \\
f_l^- (r_0) & = & \left[ A_l^+ 
\Gamma + A_l^- \right] J_{-a} (k r_0) ,\nonumber \\
g_l^- (r_0) & = & \left[ A_l^+  \Gamma C_- -
A_l^- C_+ \right] J_{-a+1} (k r_0), \nonumber
\end{eqnarray}
and, for $1/2<a<1$,
\begin{eqnarray}
f_l^+ (r_0) & = & \left[ B_l^+ + \Gamma B_l^- \right] J_{a} (k r_0) ,
\nonumber \\
g_l^+ (r_0) & = & 
\left[ B_l^+ C_- - \Gamma B_l^- C_+ \right] J_{a-1} (k r_0) ,\nonumber \\
f_l^- (r_0) & = & \left[ B_l^+ \Gamma + B_l^- \right] J_{a} (k r_0) ,
\nonumber \\
g_l^- (r_0) & = & \left[ - B_l^+  \Gamma C_- +B_l^- C_+ \right] J_{a-1} 
(k r_0). \nonumber
\end{eqnarray}
Considering the above condition (\ref{eq:4}) this imposes 
\begin{equation}
\Theta^{\pm} = \left\{
\begin{array}{ll}
+\frac{\pi}2 & 0<a<\frac12 \\
-\frac{\pi}2 & \frac12 < a< 1
\end{array}
\right. \label{thetaI}.
\end{equation}
We refer to this choice as I. Our second choice, called II, corresponds to
\beq
\Theta ^ {\pm} = \left\{ 
\begin{array}{rr}
-\frac \pi 2 & \mbox{for }\kappa>0 \\
\frac \pi 2 & \mbox{for }\kappa<0 
\end{array}
\right. .
\eeq
This extension has been considered in \cite{hage90-64-503} and it arises 
when a finite radial flux, which is ultimately shrunk to zero, is considered. 

\bigskip

We are now in a position to impose MIT boundary conditions at the exterior
boundary $r=R$. These boundary conditions ensure that the fermion current 
across the boundary vanishes. The relevant boundary operator is 
\begin{equation}
B = \mathrm{1} - i \not\!n = \left( 
\begin{array}{cc}B_+ & 0 \\ 0 & B_- \end{array} \right),
\end{equation}
with
\begin{equation}
B_\pm = \left( \begin{array}{cc}1 & \pm i e^{-i \theta} \\
\mp i e^{i \theta} & 1
\end{array} \right),
\end{equation}
and the boundary conditions reads 
\beq
B \Psi_E (R,\theta, z)=0.\nn
\eeq
Given the previous discussion for type I, see 
Eqs.~(\ref{eq:4}) and (\ref{thetaI}), 
this boundary condition has to be applied on 
the following set of spinors,
\beq
n\ge l+1: & &\quad \psi_e^{(n,1)},\,\,\psi_e^{(n,3)},\nn\\
n\le l-1: & &\quad \psi_e^{(n,2)},\,\,\psi_e^{(n,4)},\nn\\
n=l : & &\left\{\begin{array}{ll}
0<a<1/2 & \psi_e^{(n,1)},\,\,\psi_e^{(n,3)} \\
1/2<a<1 & \psi_e^{(n,2)},\,\,\psi_e^{(n,4)}
\end{array}\right. .
\eeq
and similarly for type II. 
As a result, the eigenvalues take the form
\begin{equation}
E= \frac1R \sqrt{(k_z R)^2 + z^2 +x^2}
\end{equation}
where $z=mR$ and $x$ are the solutions of the equations obtained from
\begin{equation}
J_{\mu}^2 (x) - J_{\mu-1}^2 (x) - \frac{2 z}{x} J_{\mu}(x) J_{\mu-1}(x) =0,
\label{ec-maestra}
\end{equation}
with $\mu=\nu\pm \alpha$, $\nu = n+1/2 = 3/2, 5/2, \dots , 
\infty$ and $\alpha=a-1/2$.
The critical subspace $\nu=1/2$ 
belongs to the case $\mu=\nu-\alpha$ for $-1/2<\alpha<0$ and to the case
$\mu=\nu+\alpha$ when $0<\alpha<1/2$.
All eigenvalues have degeneracy two coming from particle and antiparticle 
states.

\section{The Casimir energy for the interior}

Now everything is prepared for the calculation of the Casimir energy in 
the zeta function regularization scheme. 
Our presentation shows only the Type I case. We comment on the 
the minor changes needed for Type II where appropriate. 
Given the translational invariance
in the $z$-direction, we define the Casimir energy density by 
\begin{equation}
E_{\mathrm{C}}^{\mathrm (int)} = - 
\frac M 2    M^{2s}
\zeta_{\mathrm int} (s)  \left|_{s=-1/2}\right. ,\label{eq:7}
\end{equation}
where the zeta function has the structure
\beq
\zeta_{\mathrm int} (s) &=& 2 R^{2s-1} 
\sum_{x} \int_{-\infty}^\infty \frac{\ud y}{2\pi} \, (y^2 + z^2 + x^2)^{-s}
\nn\\
& =& R^{2s-1}   \frac{ 
\Gamma (s-1/2)}{\sqrt{\pi} \Gamma (s)}
\sum_x \left( z^2+
x^2 \right)^{-(s-1/2)}.
\eeq
Expressed as a sum over the zeros of Eq.~(\ref{ec-maestra}), we can write
\begin{equation}
\zeta_{\mathrm int} (s) = \sum_\mu \zeta_\mu^{\mathrm (int)} (s)
\end{equation}
with the partial zeta functions
\beq
\zeta_\mu^{\mathrm (int)} (s) = R^{2s-1} 
\frac{\Gamma (s-1/2)}
{\sqrt{\pi}\Gamma (s)}
\sum_{m=1}^{\infty} \left( z^2 + x_{\mu,m}^2 \right)^{-(s-1/2)} .
\eeq
Slightly more explicit is the form 
\begin{equation}
\zeta_{\mathrm int} (s) = - \zeta_{1/2 - |\alpha|}^{\mathrm (int)}
(s) + \sum_{\nu = 1/2}^{\infty} 
\left[ \zeta_{\nu+\alpha}^{\mathrm (int)} (s) +
\zeta_{\nu-\alpha}^{\mathrm (int)} (s) \right] \label{eq:8}
\end{equation}
showing the symmetry under the transformation $\alpha\rightarrow -\alpha$. So it is 
sufficient to consider $0<\alpha <1/2$. Eq.~(\ref{eq:8}) shows furthermore
that the Casimir energy is independent of the integer part of the flux.

The analysis of the Casimir energy proceeds using the methods described in 
detail in \cite{bord96-37-895,eliz93-26-2409,bord96-182-371,bene00-61-085019}
and we follow their procedure. The starting point is the contour integral 
representation
\begin{equation}
\zeta_\mu^{\mathrm (int)} (s) = 
R^{2s-1} \frac{\Gamma (s-1/2)}
{\sqrt{\pi}\Gamma (s)} 
\int_\Gamma \frac{\ud x}{2 \pi i} 
\left( x^2 + z^2\right)^{-(s-1/2)} \frac{\ud}{\ud x}
\log \bar{F} (\mu,x)
\end{equation}
where $\bar{F} (\mu,x)$ describes the eigenvalue equation,
\begin{equation}
\bar{F} (\mu,x) = J_{\mu-1}^2 (x) - J_\mu^2 (x) + 
\frac{2 z}{x} J_\mu (x) J_{\mu -1} (x)  ,
\end{equation}
and the integration path $\Gamma$ encloses all the zeros of $\bar{F} (\mu,x)$.

Shifting the contour to the imaginary axis, one obtains the following 
representation, valid for $1/2<\Re (s-1/2) <1$,
\begin{equation}
\zeta_\mu^{\mathrm (int)} (s) =  
R^{2s-1} \frac{\Gamma (s-1/2)}
{\sqrt{\pi}
\Gamma (s)} \frac{\sin \pi(s-1/2)}
{\pi} \int_z^\infty \ud u \left( u^2 - z^2 \right)^{-(s-1/2)}
 T_{\mathrm int}(z; \mu,u)   ,
\end{equation}
where $T_{\mathrm int}(z; \mu,u)$ contains the eigenvalue  equation 
transformed to the imaginary axis,
\begin{equation}
T_{\mathrm int}(z;\mu,u)=   
\frac{\ud}{\ud u} \log \left( u^{-2(\mu-1)} F(\mu,u) \right)
\end{equation}
with
\beq
F(\mu,u) = I_\mu^2 (u) + I_{\mu-1}^2 (u) + 
\frac{2 z}{u} I_\mu (u) I_{\mu -1}(u) .\label{eq:9}
\eeq
Adding and subtracting the uniform asymptotic Debye expansion of $F(\mu ,u)$, see \cite{bord96-37-895,bene00-61-085019} for details,
one arrives at
\beq
\zeta_{\mathrm int} (s) &=& -\zeta_{1/2-|\alpha|} ^{\mathrm (int)} (s) + 
Z_{\mathrm int} (s) +\sum_{i=-1}^N A_i ^{\mathrm (int)} (s) ,\label{eq:9a}
\eeq
where $Z_{\mathrm int} (s) $ is the zeta function with the asymptotic 
terms subtracted.
With $t=\nu/\sqrt{\nu^2 + x^2}$, it reads
\begin{eqnarray}
\lefteqn{ Z_{\mathrm int}(s) = } \label{eq:9b} \\
& & R^{2s-1} \frac 1 {\sqrt{\pi} \Gamma (s) \Gamma (3/2-s)} 
\sum_{\nu=1/2,3/2, \dots}^\infty \int_z^{\infty} \ud x \,
\left(x^2 - z^2 \right)^{1/2-s}
 \left\{ T_{\mathrm int}(z; \nu+\alpha,x) + 
T_{\mathrm int}(z; \nu - \alpha,x) - 
\Delta_{\mathrm int}^{(\mathrm{N})} (\nu,x, t, z) \right\} .
\end{eqnarray}
On the other hand, the asymptotic terms are,
\begin{equation}
A_i^{\mathrm (int)} (s) = \frac{R^{2s-1}}{\pi} 
\frac{\Gamma \left(\frac{1}{2}\right)}
{\Gamma (s)\Gamma (3/2-s)}  
\sum_{\nu=1/2,3/2, \dots}^\infty \int_z^\infty \ud x \,
\left( x^2 - z^2 \right)^{-(s-1/2)} \Delta_i^{\mathrm (int)}\label{eq:9c}
\end{equation}
with
\beq
\Delta_{-1}^{\mathrm (int)} = \frac{4x}{\nu} \frac{t}{1+t} ,
\quad \Delta_0^{\mathrm (int)} = \frac{2x}{\nu^2} \frac{t^2}{1+t},
\quad
\Delta_i^{\mathrm (int)} = \nu^{-i} \frac{\ud}{\ud x} 
\sum_{j=0}^{2 i} b_{(i,j)}^{\mathrm (int)} t^{i+j}.\label{eq:9d}
\eeq
The last equation may be seen as the definition of the coefficients 
$b_{(i,j)}^{\mathrm (int)}$ and they are easily determined by a simple
computer program. Most of the $b_{(i,j)}^{\mathrm (int)}$ coefficients needed are listed in Appendix A of \cite{bene00-61-085019}. 

In order to ensure $Z_{\mathrm int} (s)$ is finite one has to subtract  
terms up to $N=3$ at least. To improve the numerical convergence we have chosen 
$N=5$ in the analysis presented later on. 

The asymptotic terms have to be analysed further in order to extract the poles 
of the theory. They are conveniently expressed in terms of the functions 
\cite{eliz98-31-1743,bene00-61-085019} 
\beq
f(s;a,b,x) = \sum_{\nu=1/2}^\infty \nu^a \left[
1+ \left( \frac {\nu } x \right) ^2 \right]^{-s-b} ,\nn
\eeq
which naturally arise after performing the $x$-integrals. In detail 
the full list of useful results is 
\begin{eqnarray}
A_{-1}^{\mathrm (int)} (s) & = &
\frac{R^{2s-1}}{\pi} \frac1{s-1} z^{2(1-s)} \int_0^1 \frac{\ud y}{\sqrt{y}} f(s;0,-1;z\sqrt{y})  , 
\\
A_0^{\mathrm (int)} (s) & = & \frac{R^{2s-1}}{\pi} z^{-2s} \int_0^1 \frac{\ud y}{y^{3/2}} f(s;1,0;z \sqrt{y}) ,
\\
A_i^{\mathrm (int)} (s) & = & \sum_{j=0}^{2 i} b_{(i,j)}^{\mathrm (int)} A_{i,j} (s)   ,
\end{eqnarray}
where
\begin{equation}
A_{i,j}^{\mathrm (int)}(s) = - R^{2s-1} 
\frac{ \Gamma \left( s + \frac{i+j-1}{2}
\right)}{\sqrt{\pi}\Gamma (s) 
\Gamma \left( \frac{i+j}{2} \right)} z^{-2s - (i+j-1)}
f\left( s;j,\frac{i+j-1}2, z\right).
\end{equation}
These representations are very suitable for a numerical analysis of the 
Casimir energy as a function of the mass and the flux.

For small masses, including vanishing mass, as well as for reading 
off the poles,
the expansions in powers of $z$ are more suitable. Again, the techniques
to obtain these have been described in \cite{bord96-37-895}. In the present
context we find 
\begin{eqnarray}
A_{-1}^{\mathrm (int)}(s) & = & 
\frac{R^{2s-1}}{\pi} \sum_{n=0}^{\infty} (-1)^n 
\frac{\Gamma (n+s-1)}{\Gamma(n+1) \Gamma (s)} \,\, 
\frac1{n+s-1/2} z^{2n} \zeta_{\mathrm H} \left(2n+2s-2,\frac12 \right) , 
\label{eq:10}\\
A_{0}^{\mathrm (int)}(s) & = & 
\frac{R^{2s-1}}{\pi} \sum_{n=0}^{\infty} (-1)^n 
\frac{\Gamma (n+s)}{\Gamma(n+1) \Gamma (s)} \,\, \frac1{n+s-1/2} z^{2n} 
\zeta_{\mathrm H} \left(2n+2s-1,\frac12 \right) , \label{eq:11}\\
A_{(i,j)}^{\mathrm (int)}(s) & = & \frac{R^{2s-1}}{\sqrt{\pi}} 
\sum_{n=0}^{\infty} (-1)^{n+1} \frac{\Gamma \left(n+s + 
\frac12 (i+j-1)\right)}
{\Gamma(n+1) \Gamma (s) \Gamma \left(\frac12 (i+j)\right)} z^{2n} 
\zeta_{\mathrm H} \left(2n+2s+i-1,\frac12 \right),\label{eq:12} 
\end{eqnarray}
which makes the pole structure very explicit. 

Before studying this, let us
indicate the treatment of the `extra' contribution in Eq.~(\ref{eq:8}), $-\zeta_{1/2-|\alpha|}^{\mathrm (int)}(s)$. Given no $\nu$-sum is involved,
it is actually sufficient to just subtract the large argument expansion 
of $T_{\mathrm int} (z;1/2-|\alpha | ,x)$. The leading terms in this expansion are
\beq
T_{\mathrm int} (z;1/2-|\alpha | ,x) \sim 2+\frac{2|\alpha |} x +\frac{\alpha ^2 -z}
{x^2} +\frac{z^2} {x^3} +{\cal O} (x^{-4}) ,\label{crit:1}
\eeq
and a suitable representation to find the analytical continuation in 
the critical subspace is 
\beq
\zeta_{1/2-|\alpha |}^{\mathrm (int)} (s) &=& R^{2s-1} \frac{1} {\sqrt{\pi} 
\Gamma (s) \Gamma (3/2-s)} \times \nn\\
& &\left\{ \int_z^1 dx \,\, (x^2-z^2) ^{1/2-s} T_{\mathrm int} (z;1/2-|\alpha | ,x)
\right.\nn\\
& & + \int_1 ^\infty  dx \,\, (x^2-z^2) ^{1/2-s} \left[
T_{\mathrm int} (z;1/2-|\alpha | ,x)
-2-\frac{2|\alpha |} x-\frac{\alpha ^2 -z}
{x^2} -\frac{z^2} {x^3}\right] \nn\\
& &\left.+\int_1 ^\infty  dx \,\, (x^2-z^2) ^{1/2-s} \left[
2+\frac{2|\alpha |} x +\frac{\alpha ^2 -z}
{x^2} +\frac{z^2} {x^3}\right] \right\} .\label{eq:13}
\eeq
While the first two terms are suitable for a numerical evaluation
at $s=-1/2$,
the asymptotic terms are standard representations of hypergeometric functions
\cite{grad65b} .

We are now fully prepared to analyse the pole structure and, if a physically 
senseful interpretation of finite parts is possible, also for a numerical 
analysis of the Casimir energy. As mentioned, the poles are best read off 
in the representations (\ref{eq:10})---(\ref{eq:12}) and 
(\ref{eq:13}). The residues of the
single asymptotic terms are 
\begin{eqnarray}
\left. \mathrm{Res} \right|_{s\rightarrow -\frac12} A_{-1}^{\mathrm (int)} (s) & = &\frac1{R^2}
\left[ - \frac{z^4}{8 \pi} - \frac{z^2}{24 \pi} \right],
\\
\left. \mathrm{Res} \right|_{s\rightarrow -\frac12} A_{0}^{\mathrm (int)} (s) & = & 0 ,
\\
\left. \mathrm{Res} \right|_{s\rightarrow -\frac12} A_{1}^{\mathrm (int)} (s) & = & \frac1{R^2} \left[
- \frac{z^3}{2 \pi} + \frac{z^2}{12 \pi} + \frac{z^2 \alpha^2}{2 \pi} \right] ,
\\
\left. \mathrm{Res} \right|_{s\rightarrow -\frac12} A_{2}^{\mathrm (int)} (s) & = & 0 ,
\\
\left. \mathrm{Res} \right|_{s\rightarrow -\frac12} A_{3}^{\mathrm (int)} (s) & = & \frac1{R^2} \left[
\frac{z^3}{3 \pi} - \frac{z^2}{12 \pi} - \frac{z}{30 \pi} + \frac1{126 \pi} \right],
\end{eqnarray}
the extra contribution adds 
\beq
\left. \mathrm{Res} \right|_{s\rightarrow -\frac12} -\zeta _{1/2-|\alpha|} 
^{\mathrm (int)} (s) = -\frac{z^2} {2\pi R^2} \left( |\alpha | -\frac 1 2 \right) .
\nn
\eeq
Summing up, the total residue reads
\begin{eqnarray}
\left. \mathrm{Res} \right|_{s\rightarrow -\frac12} 
\zeta_{\mathrm int} (s) & = &\frac1{R^2} \left[
- \frac18 \frac{z^4}{\pi} - \frac{z^3}{6 \pi} + \left(\frac5{24 \pi}
+  \frac{1}{2 \pi} 
\left( \alpha^2- \left| \alpha \right| \right)\right) z^2 -
\frac1{30} \frac{z}{\pi} + \frac1{126} \frac1{\pi} \right].\nn
\eeq
So the pole  in the Casimir energy depends on the mass, as to be expected,
as well as on the flux via $\alpha$. This last dependence on $\alpha$ 
did not occur in $2+1$ dimensions, so that the difference of two situations
with different fluxes gave a finite answer \cite{bene00-61-085019}. Here
this happens only in the case $z=0$ so 
that we should restrict to this case in order
to have unambiguous answers. In the case $z\neq 0$ a flux-dependent counterterm $FR$ has to be 
introduced and finite ambiguities remain. The numerical results for the 
Casimir energy difference between the cases with and without flux for $z=0$
are presented in Fig.~\ref{fig-1} in a black line. Other results, affected by the regularization procedure, are presented as the grey lines, corresponding, from top to bottom, to the values of $z=1/64, 1/32,1/16, 1/8, 3/16$.
For this numerical analysis a minimal subtraction was considered.
\begin{figure}
\epsffile{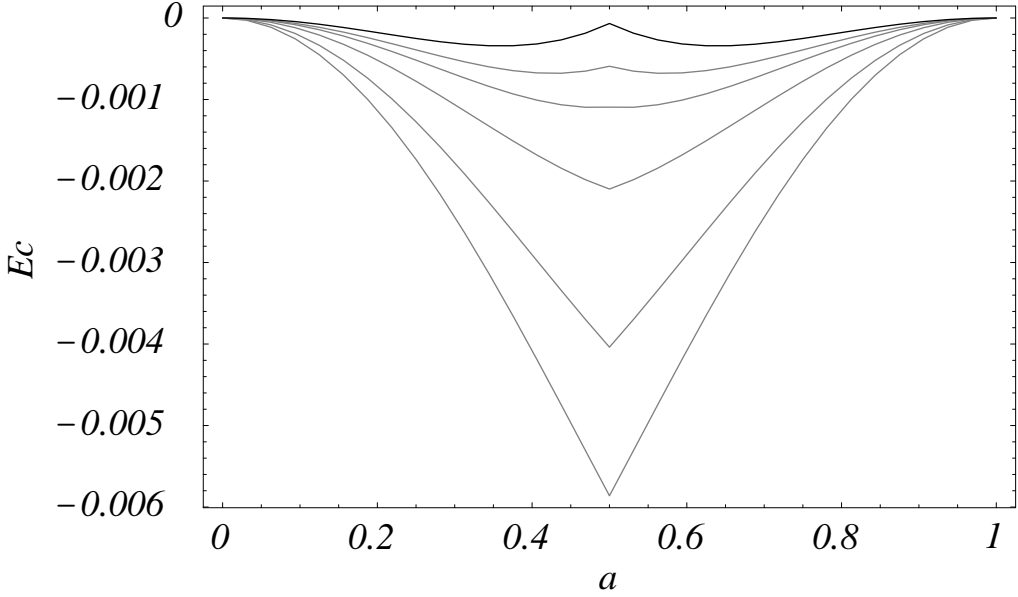} \caption{(Differences of)Interior Casimir Energies - Type I}
\label{fig-1}
\end{figure}

Our analysis can be straightforwardly extended to the Type II case. 
It is not difficult to show that just 
the replacement of $|\alpha|$ by $\alpha$ in the extra term of 
Eq.~(\ref{eq:8}), $-\zeta_{1/2-\alpha}^{\mathrm (int)} (s)$, 
or in the second 
order of the asymptotic expansion, leads to the wanted results. A 
similar replacement rule gives the total residue and all 
the considerations above apply directly. The corresponding 
numerical results are shown in Fig.~\ref{fig-2}

\begin{figure}
\epsffile{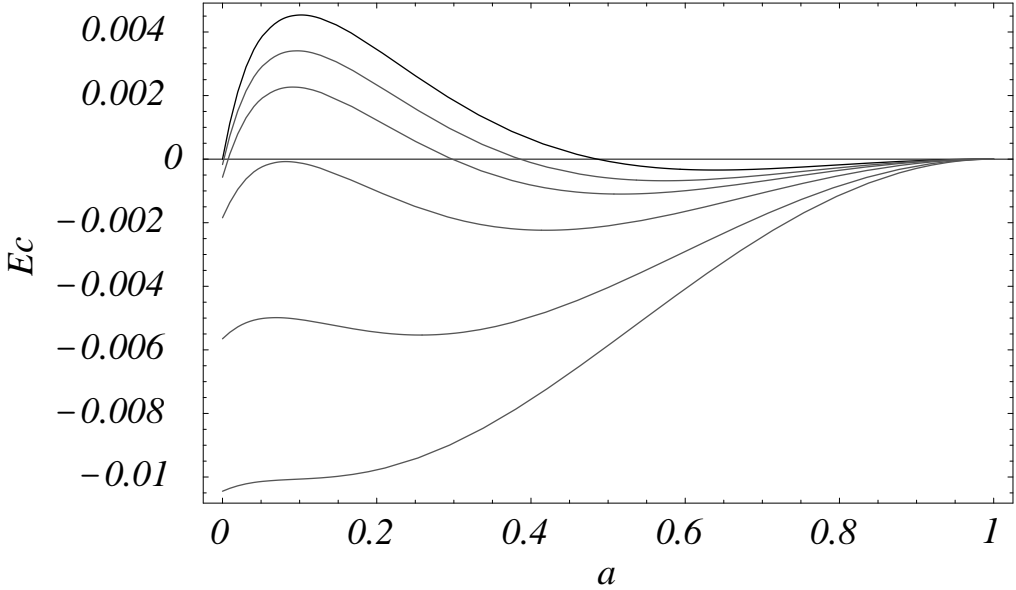} \caption{(Differences of) Interior Casimir Energies - Type II}
\label{fig-2}
\end{figure}

\section{The Casimir energy of the external sector}
Let us now consider the external sector in order to see if the pole structure
considerably simplifies when combining 
the internal and external spaces. Partly
this is to be expected for simple geometric reasons, because the extrinsic 
curvature has the opposite sign. 

The external sector is analysed most effectively using the formulation 
in terms of a Jost function, see e.g.~\cite{bord96-53-5753}. As a first step,
a large `external' sphere is introduced as an additional boundary to compactify
the space. Boundary conditions are imposed, e.g.~MIT ones, and the 
eigenfunctions are combinations of the ones given in (\ref{eq:2}) 
and (\ref{eq:3}) with the $J_\omega$ and $J_{-\omega}$ replaced by 
$H_\omega ^{(1)}$ and $H_\omega ^{(2)}$. Having the eigenfunctions at hand, 
it is not difficult to find the Jost function for the associated scattering
problem. As observed already in various examples 
\cite{eliz98-31-1743,bord97-56-4896}, the external sector is 
obtained when $I_\nu$ is replaced by $K_\nu$. This, essentially, is true also 
in the present situation apart from the fact that the external space does not
have a critical subspace because the singular flux is not sensed directly 
and no self-adjoint extension is involved. The last comment, as it turned 
out, is just equivalent to the use of $\alpha$ instead of 
$|\alpha|$ where applicable. 

As a result of the above comments, the zeta function of the external 
part reads
\begin{equation}
\zeta_{\mathrm{ext}} = 
-\zeta_{1/2-\alpha}^{\mathrm (ext)} (s) + \sum_{\nu=1/2}^{\infty} 
\left[ \zeta_{\nu+\alpha}^{\mathrm (ext)} (s) + 
\zeta_{\nu-\alpha}^{\mathrm (ext)} (s)\right] ,
\end{equation}
where
\begin{equation}
\zeta_{\mu}^{\mathrm (ext)} (s)= R^{2s-1} \frac{
1}{\sqrt{\pi}\Gamma (s) \Gamma (3/2 -s)}
 \int_z^\infty \ud x \, (x^2 -
z^2)^{-(s-1/2)}  T_{\mathrm ext}(z;\mu,u),
\end{equation}
\begin{equation}
T_{\mathrm ext}(z;\mu,u)=\frac{\ud}{\ud x} \log \left( u^{2 \mu}
G(\mu,u) \right),
\end{equation}
and
\begin{equation}
G(\mu,u) = K_\mu^2 (u) + K_{\mu-1}^2 (u) + \frac{2 z}{u} K_\mu (u)
K_{\mu -1}(u).
\end{equation}
In clear analogy the equations corresponding to (\ref{eq:9a})---(\ref{eq:9d}) 
can be written down, where one has the relations
\beq
\Delta_{-1} ^{\mathrm (ext)} = -\Delta_{-1}^{\mathrm (int)} , \quad 
\Delta_{0} ^{\mathrm (ext)} = -\Delta_{0}^{\mathrm (int)} ,\nn
\eeq
and the multipliers $b_{(i,j)}^{\mathrm (ext)}$ are obtained in a similar 
way as the internal multipliers 
$b_{(i,j)}^{\mathrm (int)}$. Also the `extra' contribution $-\zeta
_{1/2-\alpha}^{\mathrm (ext)}$ is dealt with as before, see eq.~(\ref{crit:1}) 
and (\ref{eq:13}), but 
now we have
\beq
T_{\mathrm ext} (z;1/2-\alpha  ,x) \sim -2-\frac{2\alpha } x -\frac{\alpha ^2 +z}
{x^2} +\frac{z^2} {x^3} +{\cal O} (x^{-4})  .\nn
\eeq
The residues in the exterior space are thus obtained from the interior ones
by the simple replacement rules already 
mentioned, i.e. $z\rightarrow -z$ and $|\alpha| \rightarrow \alpha$. 
As a result, the Casimir energy difference between the case with and 
without flux for massless fields is finite and is 
shown as the black line in Fig.~\ref{fig-3}. The grey 
lines show the regularization 
dependent results for, from bottom to top, $z=1/64,1/32,1/16,1/8,3/16$. 

\begin{figure}
\epsffile{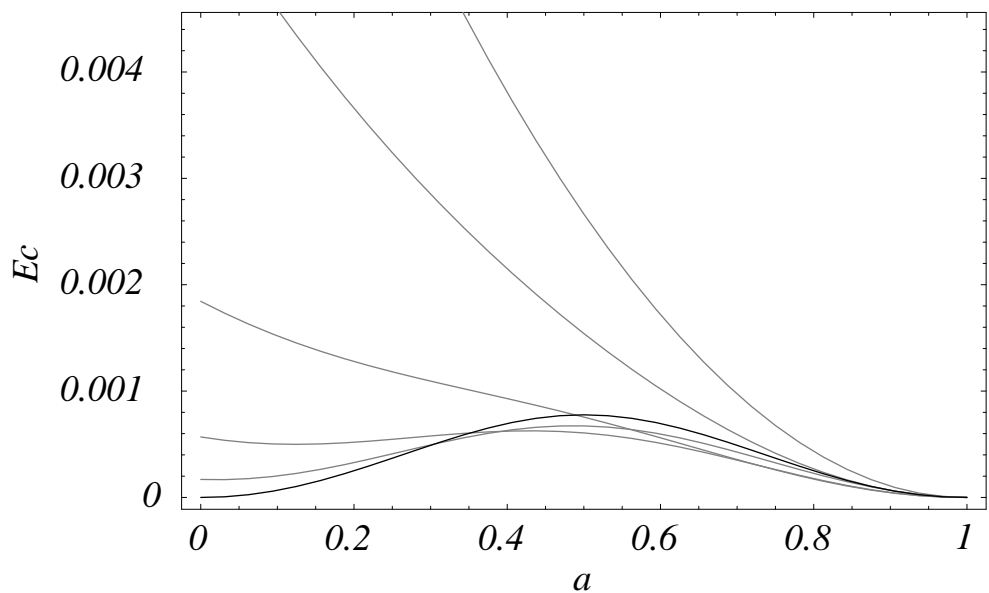} \caption{(Differences of) Exterior Casimir Energies}
\label{fig-3}
\end{figure}

\section{The Casimir energy in the whole space - Final Comments}

Adding up the interior and the external sector, we find the pole
\begin{equation}
\left. \mathrm{Res} \right|_{s\rightarrow -\frac12}
\zeta_{\mathrm int} (s)+ \left. \mathrm{Res}
\right|_{s\rightarrow -\frac12} \zeta_{\mathrm ext} (s) =
\frac1{R^2} \left[   -\frac{z^3}{3 \pi} - \frac1{15}
\frac{z}{\pi}  - \frac12 \frac{z^2}{\pi} \left( -\alpha + \left|
\alpha \right| \right)   \right]
\end{equation}
for behaviour I, still showing a flux dependent term. Numerical results for the difference between arbitrary and integer flux (only unambiguous for the black $z=0$ curve) are shown in Fig.~\ref{fig-4}.

\begin{figure}
\epsffile{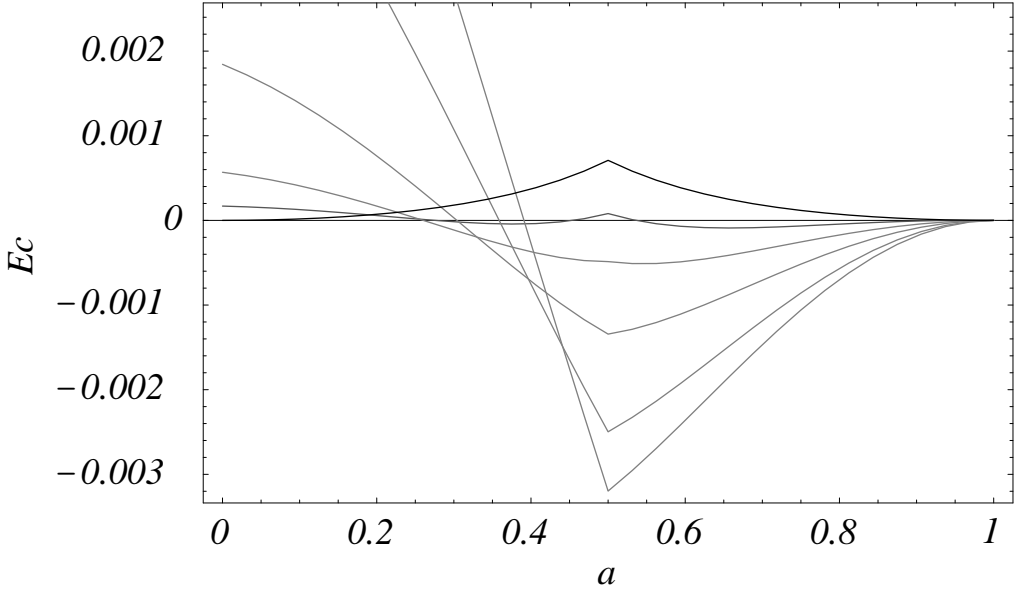} \caption{(Differences of) Complete Casimir Energies - Type I}
\label{fig-4}
\end{figure}

On the other hand, 
\begin{equation}
\left. \mathrm{Res} \right|_{s\rightarrow -\frac12}
\zeta_{\mathrm int} (s)+ \left. \mathrm{Res}
\right|_{s\rightarrow -\frac12} \zeta_{\mathrm ext} (s) =
\frac1{R^2} \left[   -\frac{z^3}{3 \pi} - \frac1{15}
\frac{z}{\pi}  
  \right]
\end{equation}
gives the residue for behaviour II. We see that the Casimir energy difference
between two flux-situations is finite and we can fully
analyse its dependence on the mass. Numerical results are displayed in 
Fig.~\ref{fig-5}, for the values of $z$ already given, from top to bottom. 
Whereas for non-vanishing mass the field free situation leads to the lowest
Casimir energy, for non-zero mass a non-trivial global minimum of the 
Casimir energy emerges. Above a certain critical flux, quantum contributions
support the stability of the flux, 
at least for type II.

\begin{figure}
\epsffile{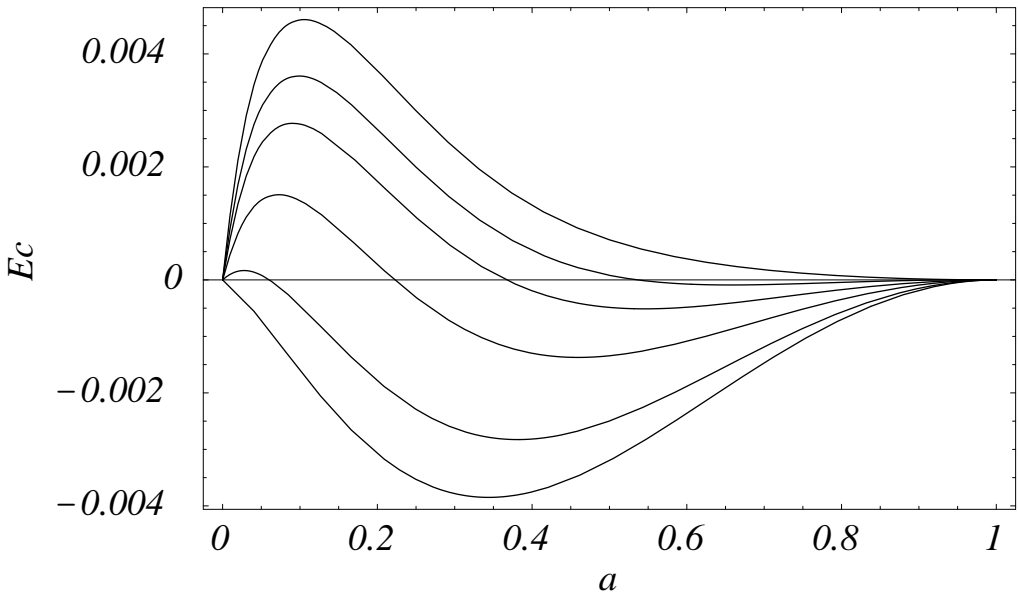} \caption{(Difference of) Casimir Energies - Type II}
\label{fig-5}
\end{figure}

In summary, we have studied in detail 
the behaviour of the Casimir energy of Dirac fields 
in cylindrically symmetric situations. 

In contrast to the $2+1$-case, when studying the interior case, 
flux-dependent residues do not vanish when just 
considering the difference between two values of the flux (except for the massless case). In our hope of obtaining finite results when considering the whole space, 
the external space was analysed too. 
The geometrically grounded expectation of a simpler pole 
structure for the whole space eventually proved to be right, at least for 
the Type II choice of the self-adjoint extension. 
Completely well-defined results for the massive 
case were obtained numerically.  

The presence of the mass and the choice of the self-adjoint 
extension for the radial Hamiltonian at the origin are 
crucial in our results which emphasizes that, for curved 
boundaries, the contribution of the mass 
is not exponentially small \cite{bord97-56-4896} as it is for parallel
plates \cite{Plunien1986}. Concerning the self-adjoint extensions, 
the remarkable differences in the resulting Casimir energies 
reinforce the well-known fact that the one-parameter 
family of self-adjoint extensions describes nontrivial 
physics in the core, see also \cite{gerb89-40-1346}. 
Furthermore, a better understanding of the physical meaning of the different
self-adjoint extensions, by considering how the Casimir energy
depends on the parameter $\Theta$ of the one-parameter family of
self-adjoint extensions, should be envisaged in the future.

\acknowledgements

We thank Stuart Dowker for interesting and helpful discussions on the subject.
KK has been supported by
the EPSRC under Grant No GR/M45726. MDF has
been supported by CONICET.

%\bibliography{biblio}

\end{document}